\documentclass{aa}
\usepackage{graphicx}
\usepackage[]{natbib}
\bibpunct{(}{)}{;}{a}{}{,} 
\usepackage{txfonts}
\newcommand{\kms}{km\,s$^{-1}$}
\begin{document}
   \title{The wideband backend at the MDSCC in Robledo}
   \subtitle{A new facility for radio astronomy at Q- and K- bands}

   \author{
   J. R. Rizzo\inst{1}
   \and
   A. Pedreira\inst{2}
   \and
   M. Guti\'errez Bustos\inst{1}
   \and
   I. Sotuela\inst{3}
   \and
   J. R. Larra\~naga\inst{2}
   \and
   L. Ojalvo\inst{2}
   \and
   M. Franco\inst{4}
   \and
   J. Cernicharo\inst{1}
   \and
   C. Garc\'{\i}a-Mir\'o\inst{3}
   \and
   J. M. Castro Cer\'on\inst{3}
   \and
   T. B. H. Kuiper\inst{4}
   \and
   M. V\'azquez\inst{3}
   \and
   J. Calvo\inst{3}
   \and
   A. Baquero\inst{2}
   }

   \institute{Centro de Astrobiolog\'{\i}a (INTA-CSIC), 
              Ctra.~M-108, km.~4, E-28850 Torrej\'on de Ardoz, Madrid, Spain\\
              \email{ricardo@cab.inta-csic.es}
   \and
              Instituto Nacional de T\'ecnica Aeroespacial,
              Ctra.~M-108, km.~4, E-28850 Torrej\'on de Ardoz, Madrid, Spain
   \and
              Madrid Deep Space Communications Complex, 
              Ctra.~M-531, km.~7, E-28294 Robledo de Chavela, Madrid, Spain
   \and
              Jet Propulsion Laboratory, California Institute of Technology, 
              4800 Oak Grove Drive, Pasadena, CA 91109
              }

   \date{Received January 18, 2012; accepted March 15, 2012}

 
  \abstract
   {The antennas of NASA's Madrid Deep Space Communications Complex (MDSCC) 
    in Robledo de Chavela are available as single-dish radio astronomical 
    facilities during a significant percentage of their operational time. Current  
    instrumentation includes two antennas of 70 and 34\,m in diameter, equipped 
    with dual-polarization receivers in K (18 -- 26\,GHz) and Q (38 -- 50\,GHz) 
    bands, respectively. Until mid-2011, the only backend available in MDSCC was 
    a single spectral autocorrelator, which provides bandwidths from 2 to 16\,MHz. 
    The limited bandwidth available with this autocorrelator seriously limited 
    the science one could carry out at Robledo.
   }
   {We have developed and built a new wideband backend for the Robledo antennas, 
    with the objectives (1) to optimize the available time and enhance 
    the efficiency of radio astronomy in MDSCC; and (2) to tackle new scientific 
    cases impossible to that were investigated with the existing autocorrelator.}
   {The features required for the new backend include (1) a broad 
    instantaneous bandwidth of at least 1.5\,GHz; (2) high-quality and stable 
    baselines, with small variations in frequency along the whole band; (3) easy 
    upgradability; and (4) usability for at least the antennas that host the K- 
    and Q-band receivers.
   }
   {The backend consists of an intermediate frequency (IF) processor, a fast Fourier 
    transform spectrometer (FFTS), and the software that interfaces and manages the 
    events among the observing program, antenna control, the IF processor, the FFTS 
    operation, and data recording. The whole system was end-to-end assembled in 
    August 2011, at the start of commissioning activities, and the results are 
    reported in this paper. Frequency tunings and line intensities are stable over 
    hours, even when using different synthesizers and IF channels; no aliasing 
    effects have been measured, and the rejection of the image sideband was 
    characterized.
   }
   {The new wideband backend fulfills the requirements and makes better use 
    of the available time for radio astronomy, which opens new possibilities to 
    potential users. The first setup provides 1.5\,GHz of instantaneous bandwidth in 
    a single polarization, using 8192 channels and a frequency resolution of 212\,kHz; 
    upgrades under way include a second FFTS card, and two high-resolution cores 
    providing 100\,MHz and 500\,MHz of bandwidth, and 16384 channels. These upgrades 
    will permit simultaneous observations of the two polarizations with instantaneous 
    bandwidths from 100\,MHz to 3\,GHz, and spectral resolutions from 7 to 212\,kHz.
   }


   \maketitle
%

\section{Radio astronomy activity in Robledo}

The NASA Deep Space Network\footnote{See {\tt http://deepspace.jpl.nasa.gov}} 
(DSN) is the international network of antennas used to support tracking and 
operations activities of space missions usually beyond the Earth orbits. The DSN 
currently consists of three complexes located at Goldstone (USA), Canberra 
(Australia), and Robledo de Chavela (Spain). Each complex hosts several 
sensitive, large-aperture antennas, whose diameters are between 26 and 70\,m.

The Robledo complex, namely Madrid Deep Space Communications Complex (MDSCC), 
has six of these antennas. A fraction of the operational time of the MDSCC 
antennas are routinely used for radio astronomical observations. In addition to 
its use as a VLBI station through the European VLBI Network (EVN), the antennas 
are usable as single-dish radio telescopes by means of two different programs, 
the ``Guest Observer'' program, which involves all the DSN antennas and is 
managed directly by JPL, and the so-called ``Host Country'' radio astronomy 
program, which is a guaranteed time program for Spanish astronomers managed by 
INTA (Instituto Nacional de T\'ecnica Aeroespacial). All six antennas and their 
corresponding receivers are available, although two instruments are actively 
used for radio astronomy: the DSS-63 antenna, 70\,m in diameter, equipped with a 
K-band (18 -- 26\,GHz) receiver; and the DSS-54 antenna, 34\,m in diameter, 
equipped with a Q-band (38 -- 50\,GHz) receiver. Table 1 summarizes the main 
technical characteristics of these two antennas.

%
\begin{table}
\caption{The two major antennas at MDSCC, Robledo}
\label{table:1}      
\begin{center}
\begin{tabular}{l@{\hspace{2em}}r@{\hspace{1.5em}}r}
\hline\hline
Antenna & DSS-63 & DSS-54\\
Diameter (m)& 70 & 34\\
Type$^1$ & HEF & BWG\\
Latitude$^2$ (deg)  &  40:25:52 &  40:25:32\\
Longitude$^2$ (deg) & 355:45:71 & 355:44:45\\
Elevation$^2$ (m)   & 865 & 838\\
Receiver band & K & Q\\
Frequency range (GHz) & 18 -- 26 & 38 -- 50\\
Half-power beam width ($"$) & 42 & 43\\
Receiver temperature (K) & $25$ & $40$\\
System temperature (K) & 45 -- 90 & 80 -- 100\\
Typical opacity at zenith & 0.08 & 0.10\\
Aperture efficiency & 0.50 & 0.46\\
Sensitivity (Jy/K) & $2.8$ & $6.7$\\
\hline
\end{tabular}
\end{center}
\noindent $^1$ HEF: High-efficiency Cassegrain focus. BWG: Beam waveguide, 
               Coud\'e focus.\\
\noindent $^2$ Geodetic coordinates in the ITRF93 reference system.
\end{table}
%

The two bands mentioned in Table 1 include several important molecular lines, 
and the angular resolution of both antennas are similar (slightly above $40"$, 
depending on the observing frequency). Single-dish radio astronomy at MDSCC can 
address a wide variety of scientific categories, such as the 
studies of the interstellar and circumstellar media, star formation and 
evolution, and solar system bodies \citep{gom06, sua07, bus09, hor10}. Until mid 
2011, the only backend available at MDSCC was an autocorrelator, which provides 
excellent spectral resolution, (as low as 5\,kHz) but instantaneous bandwidths 
of 2, 4, 8, and 16 MHz. This limitation seriously affected the possible science 
cases; there are no practical possibilities to perform, for example, spectral 
surveys, simultaneous line observations, extragalactic astronomy, or the 
observations of a single spectral line in sources that have a wide velocity 
range of emission.

Aiming to overcome the bandwidth limitation problem, we have started in 2009 the 
design and construction of a new complete backend system for its use in the 
Robledo antennas. The backend was end-to-end assembled by mid-2011, when the 
commissioning tests of the first setup started. This paper reports the main 
features of the new backend, some astronomical results of the commissioning, 
and prospects for future upgrades.


\begin{figure*}
  \sidecaption
  \includegraphics[width=12cm]{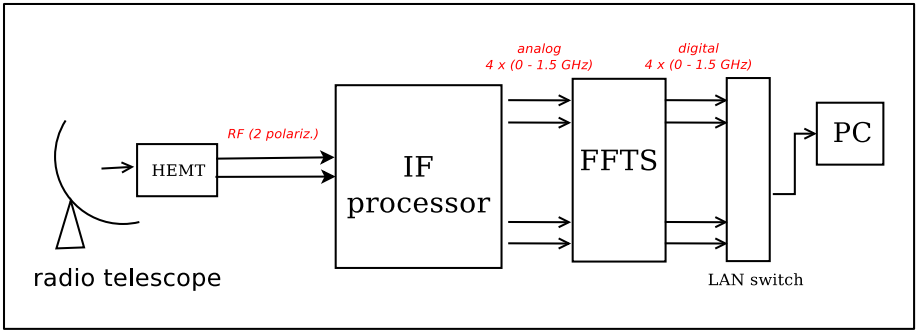}
  \caption{
Concept diagram of the signal path from the antenna receiver (HEMT) through 
the backend. The first stage of the backend is the IF processor, followed by 
the FFT spectrometer. Finally, data are processed and recorded on a dedicated 
PC (see text).
  }
\end{figure*}

\begin{figure*}
\centering
  \includegraphics[width=\textwidth,angle=0]{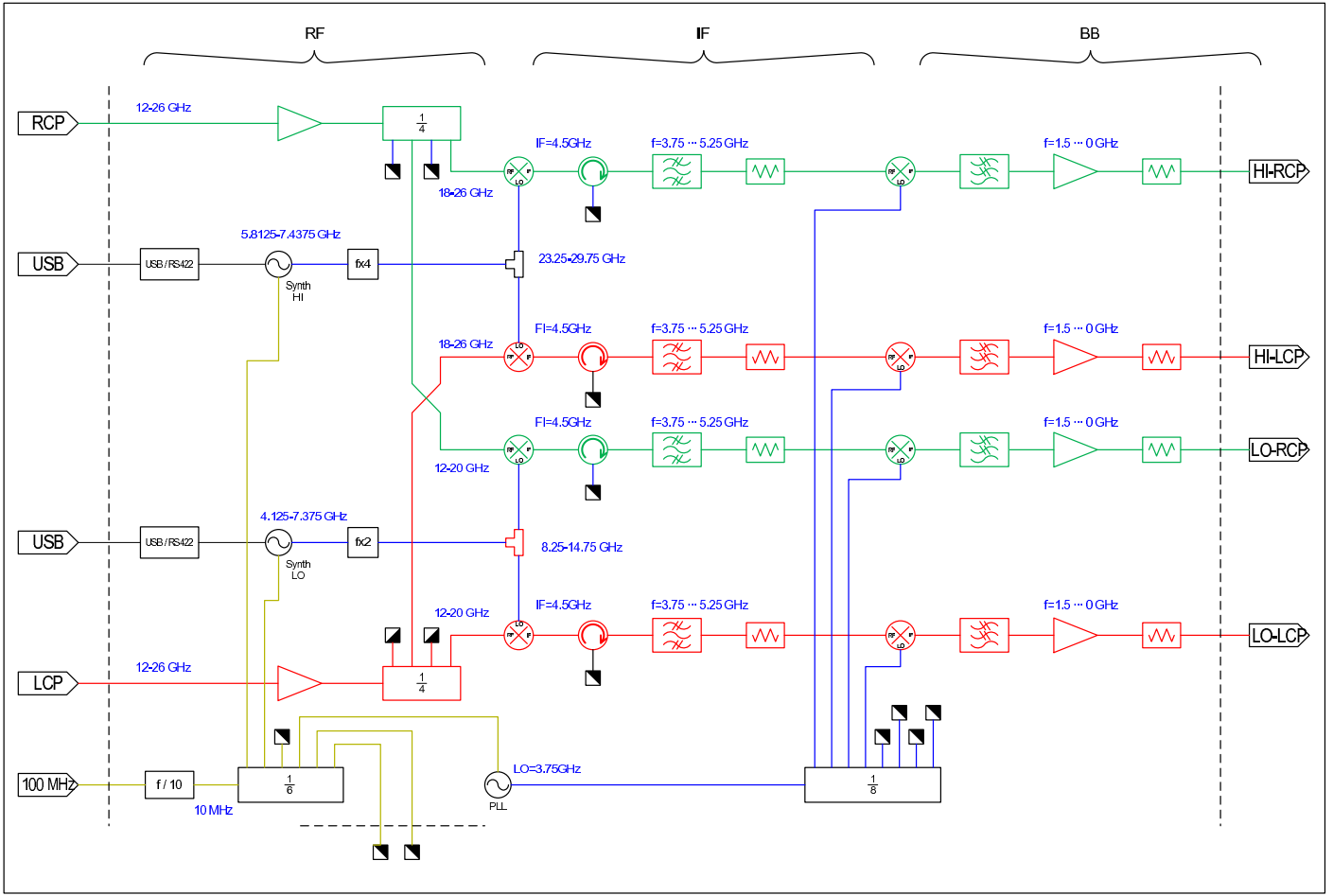}
  \caption{
Block diagram of the IF processor. Input polarizations are marked by red and 
green colors. The signals are divided into two channels, LO and HI, for each 
polarization, providing four output channels at base band.
  }
\end{figure*}


\section{The new wideband backend}
\subsection{Aims and features}

The new backend was conceived with the objectives to optimize the use of available 
time for single-dish radio astronomy in MDSCC and to tackle new scientific cases, 
which was impossible to do using the current autocorrelator. The current high 
sampling rate of commercially available analog-to-digital converters (ADCs) 
together with the high power of field programable gate array (FPGA) chips allowed 
us to consider a complete backend able to process several GHz of instantaneous 
bandwidth. Large bandwidth spectrometers have been developed at several radio 
telescopes since the early 2000s. Also, high-performance electronics are now 
commercially available, ensuring reliable intermediate frequency (IF) processing, 
maintaining the quality of such large bandwidths. 

Two other features are required for the backend. First, it has to be easily 
upgradeable to permit enhancements without the need to discard already 
acquired/built components. Second, the new backend has to be useful for the two 
major receivers that operate in single-dish radio astronomy at the MDSCC (the Q- 
and K-band receivers already mentioned).

Fig.~1 sketches the concept of the backend. The fist stage is the radio signal 
reception, by means of the K- or Q-band receivers. The K-band frequency range 
goes from 18 to 26\,GHz, while the Q-band receiver is sensitive from 38 to 
50\,GHz; the Q-band signal is mixed to a local oscillator at 62\,GHz within the 
receiver, which results in a mirrored band transport to the range 12 to 24\,GHz. 
Therefore, a backend usable for both receivers should be able to process signals 
from 12 to 26\,GHz in two polarizations. The two-polarization radio signal (RF 
in Fig.~1) goes then to the first part of the backend, namely the IF processor. 
The IF processor splits each polarization radio signal into two channels and 
downconverts the resulting four channels, giving four base band (BB) analog 
outputs from 0 to 1.5\,GHz. The second step of the backend is a fast Fourier 
transform spectrometer (FFTS), which digitizes and samples the BB signals. The 
final step is to manage and record the data, which is done through a program on 
a dedicated PC.

%
\begin{table}[b]
\caption{Hardware specifications of the IF processor}
\label{table:2}      
\centering
\begin{tabular}{l@{\hspace{4em}}r}
\hline\hline
Input frequency range               & 12 -- 26\,GHz\\
Number of input signals             & 2            \\
Output frequency range              & 0 -- 1.5\,GHz\\
Number of output signals            & 4\\
Intermediate frequency              & $4.5\pm0.75$\,GHz\\
Frequency range of the LO channels  & 12 -- 20\,GHz\\
Frequency range of the HI channels  & 18 -- 26\,GHz\\
Power consumption                   & $<40$\,W, $<500$\,mA\\
Supply                              & 110 or 220\,V\\
External frequency reference        & 100\,MHz\\
Synthesizers control ports          & 2 $\times$ RS-422\\
Interface to PC                     & 2 $\times$ USB 2.0\\
\hline
\end{tabular}
\end{table}
%

\subsection{The IF processor}

The IF processor was designed and built at the INTA's radar laboratory, with the 
exception of several semi-rigid cables, which have been assembled at JPL.

The Fig.~2 depicts a block diagram of the IF processor, while Table~2 summarizes 
its specifications. Two input signals are shown, which correspond to both 
polarizations of the radio signal. Each polarization is then split into two parts, 
which are meant as low- and high-frequency channels (LO and HI, respectively); the 
frequency ranges are 12 -- 20\,GHz for the LO channels, and 18 -- 26\,GHz for the 
HI channels. 

Two synthesizers (HI and LO) generate the adequate frequencies; these signals 
pass through multipliers ($\times 4$ in HI, $\times 2$ in LO) and are mixed 
to the radio signal. These operations translate the signal in an IF of 4.5\,GHz. 
The IF signal is filtered to a width of 1.5\,GHz. A second mixing is then 
produced at a fixed frequency of 3.75\,GHz, which translates the signals in 
four BB signals, 1.5\,GHz in width, as outputs. The HI channels spectra are 
mirrored as a result of the first frequency mixing.

The system works with an external reference of 100\,MHz. A divider generates 
10\,MHz internal references, which are used by the synthesizers and the local 
oscillator of the second mixing. In addition, two other 10\,MHz output references 
are provided for external purposes. The synthesizers are controlled by RS-422 
serial ports; these ports are interfaced to standard USB 2.0 by two cards, which 
allow a PC to control and tuning of the synthesizers.

\subsection{The fast Fourier transform spectrometer}

The analog BB signals are sampled and digitized by the FFTS. This spectrometer, 
based on FPGA chips Virtex-4, is essentially the same as the one developed for 
APEX \citep{kle06} and the IRAM 30-m radiotelescopes, and has been constructed 
by the same group. The spectrometer can provide a maximum instantaneous bandwidth 
of 1.5\,GHz through a four-tap polyphase filter bank algorithm \citep{kle08}.

The main features of the FFTS are summarized in Table 3. The default number of 
channels and bandwidth are 8192 and 1.5\,GHz, respectively, though other 
configurations are possible. The spectral resolution (equivalent noise bandwidth, 
ENBW) is 1.16 times the frequency spacing. 
%

\begin{table}[b]
\caption{Specifications of the fast Fourier transform spectrometer}
\label{table:3}      
\centering
\begin{tabular}{l@{\hspace{4em}}r}
\hline\hline
ADC sample rate                     & 3\,GS/s (max)\\
ADC resolution (quantisation)       & 8 bits\\
Maximum instantaneous bandwidth     & 1.5\,GHz\\
Maximum number of channels          & 8192\\
Maximum number of spectrometers     & 4 (2 available)\\
Interface to control commands       & ethernet\\
Interface to dataports              & ethernet\\
FPGA unit                           & Xilinx Virtex-4 SX-55\\
Time stamping                       & On-board, IRIG-B\\
ADC typical temperature             & $35-45$\,C\\
ADC maximum temperature             & 75\,C\\
Input level (ideal working regime)  & $-14.2$ to $-3.4$\,dBm\\
Input level (minimal working signal)& $-23.2$\,dBm\\
Input level (maximum tolerable)     & $+3.9$\,dBm\\
\hline
\end{tabular}
\tablefoot{Number of channels corresponds to the 1.5\,GHz setup, which 
           is the default.}
\end{table}
%

The spectrometer consists of a master unit and a set of FFTS boards. The master 
unit processes the commands, frequency references, synchronization, and 
IRIG-B\footnote{
Inter-Range Instrumentation Group, format B. IRIG-B is one of 
the most extended standard time codes used in electronic equipment where serial 
times are generated for correlation of data with time. See {\tt http://www.irigb.com}
} 
time stamps. The FFTS boards are the elements that host the ADCs and the FPGAs, 
and therefore are the units that perform the digitalisation and computes the power 
spectra through the FFT algorithm. We initially had just one FFTS board, which was 
used for the commissioning observations reported here. A second FFTS board is being 
installed (January 2012), and therefore we will process up to 3\,GHz of 
instantaneous bandwidth from any of the four available BB output channels of the 
IF processor. The master unit and each of the boards are controlled by independent 
ethernet connections. The commands are sent to the master unit, while the traffic 
of data are transmitted throughout the individual cards.

The FFTS is operated by AFFTS ({\it Array FFTS}) software, a multi-threaded program 
that has a set of libraries and functions to operate the boards independently and with flexibility. The communication is done through a 
set of UDP/SCPI socket commands\footnote{Available at {\tt 
http://www.mpifr-bonn.mpg.de/staff/\\dmuders/APEX/SCPI/APEX-MPI-ICD-0005-R1\_0.pdf}}.
A multi-core software has also been added to allow for other configurations. 
Using the appropriate anti-aliasing filters at the FFTS input, the software can 
manage bandwidths of 100 to 750\,MHz, with up to 16384 channels. 
Table 4 specifies the possible modes of operation; it is not planned to use the setup 
providing 750\,MHz of bandwidth.

%
\begin{table}
\caption{FFTS performances}
\centering
\begin{tabular}{ccccc}
\hline\hline
Setup & Bandwidth & Number of          & Freq.~spacing & ENBW \\
      &   (MHz)   & channels           &     (kHz)         & (kHz)\\
\hline
1     & 1500      & 8192               & 183               & 212\\
2     &  750      & 16384              & 46                & 53\\
3     &  500      & 16384              & 31                & 35\\
4     &  100      & 16384              & 6                 & 7\\
\hline
\end{tabular}
\tablefoot{750-MHz setup not planned to be used.}
\end{table}
%
%


\begin{figure}[b]
  \centering
     \includegraphics[width=\columnwidth]{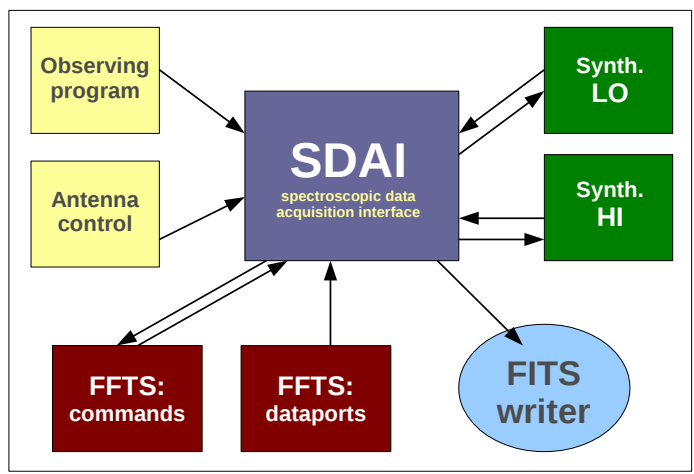}
  \caption
   {Schematic diagram of operations made by SDAI. The communication with the 
    observing program and antenna control (yellow boxes), as well as the FFTS 
    (red boxes), are made through sockets. Tuning is performed by a continuous 
    setting of the synthesizers through USB connections. Resulting spectra are 
    written in FITS format for the subsequent analysis.}
\end{figure}

\subsection{The control software}

The operation of the new backend in coordination with the existing devices 
(antennas, receivers, calibration) required a totally new software development. 
This new software acts as an interface among the different processes, and is in 
charge of the synchronization and of the data writing. The software was named 
{\it Spectroscopic Data Acquisition Interface} (SDAI hereafter). It is written 
in python 2.5, using PyQt4 for the user interface. Other libraries are used for 
FITS managing, data manipulation, and communication with the hardware and other 
programs.

A schematic diagram of SDAI operations is depicted in Fig.~3. The program 
receives astronomical information (source, frequencies, Doppler correction, 
etc.) from the observing program, and also information about the antenna 
status. These operations, labeled in yellow in Fig.~3, are made with socket 
connections. SDAI also interacts with the IF processor (green boxes in Fig.~3) 
by properly tuning its synthesizers through their USB ports; the tunings are 
made at the beginning of each scan to keep the Doppler-corrected frequencies 
at the center of the bands. The control of the FFTS (red boxes in Fig.~3) is 
performed by means of ethernet socket connections; it consists of the setup 
and the injection of commands to proceed to the integrations, and the 
subsequent management of the data. Finally, SDAI produces a FITS file for each 
FFTS board, and for each phase (source of reference).

The backend is meant to work in position-switching observing mode. Therefore, 
post-processing is required to subtract the sky contribution and calibrate 
the spectra. This is performed by an almost-real-time script (maximum delay 
of 5\,sec), which processes source-reference pairs and writes the calibrated 
spectra into a CLASS\footnote{CLASS is part of the GILDAS software, developed 
by IRAM. See {\tt http://www.iram.fr/IRAMFR/GILDAS}} file; the script was 
written in SIC scripting language using the SIC python binding.

\section{Commissioning results}

%
\begin{table*}
\caption{Summary of observations}
\centering
\begin{tabular}{lcrcrrc}
\hline\hline
\multicolumn{1}{c}{Source} & RA      & \multicolumn{1}{c}{Dec}        & Freq.~range & \multicolumn{1}{c}{$t_{\rm int}$} & \multicolumn{1}{c}{Nr.~of} & Ref.\\
       & \multicolumn{2}{c}{(J2000)} & (GHz) & (min)         & \multicolumn{1}{c}{lines} \\
\hline
WX\,Psc       & 01 06 26.0 &  12 35 53 & 42.0 -- 43.9 & 19 &  4 & 1\\
W3(OH)        & 02 27 03.9 &  61 52 26 & 47.6 -- 49.8 & 11 &  1 & 2\\
TMC-1(HC$_5$N)& 04 41 41.9 &  25 41 27 & 37.5 -- 39.0 &  6 &  3 & 3\\
              &            &           & 39.3 -- 40.8 & 19 &  2 & 4\\
              &            &           & 42.1 -- 43.6 &  5 &  2 & 4\\
              &            &           & 44.5 -- 46.1 & 22 &  3 & 5\\
TMC-1         & 04 41 45.9 &  25 41 27 & 38.3 -- 46.8 & 39 & 12 & 6\\
Orion-KL      & 05 35 14.1 & $-$05 22 22 & 42.4 -- 45.0 & 45 & 12 & 7\\
Sgr\,B2(M)    & 17 47 20.5 & $-$28 23 06 & 42.0 -- 43.5 & 24 & 16 & 8\\
W51D          & 19 23 39.8 &  14 31 10 & 42.0 -- 43.9 & 46 & 10\\
DR21-W        & 20 37 07.6 &  42 08 46 & 43.4 -- 44.9 & 15 &  1 & 9\\
\hline
\end{tabular}
\tablebib{(1)~\citet{kim10}; (2)~\citet{tur73}; (3)~\citet{tak98}; 
(4)~\citet{tak90}; (5)~\citet{kaw92}; 
(6)~\citet{kai04}; (7)~\citet{god09}; (8)~\citet{deg06}; (9)~\citet{has90}
}
\end{table*}
%
%

Since early 2010, we have performed a set of different astronomical 
observations aimed to test 
available parts of the new backend. The FFTS was tested using the 
old downconverter at 321.4\,MHz, which provided a bandwidth of 70\,MHz. Previous 
releases of SDAI were step-by-step tested after adding functionality. At the 
same time, several laboratory tests and measurements of the electronic components 
of the IF processor were made; once assembled, each IF channel was also measured. 
The stability of the FFTS has been already tested by the manufacturer through 
Allan variance measurements, which showed that it is stable within 
$\sim 1000$\,sec \citep{kle06, kle08}.

The astronomical results reported here are the first using the whole fully 
assembled system. The observations were performed aimed to test:

\begin{itemize}
\item{the performance of the whole backend, particularly the different IF channels}
\item{the quality and stability of tunings with different synthesizers}
\item{the band stability during long integrations}
\item{the performances of the different filters}
\item{the image sideband reflections, spurious signals, spikes, aliasing.}
\end{itemize}


\begin{figure}
\centering
  \includegraphics[width=\columnwidth]{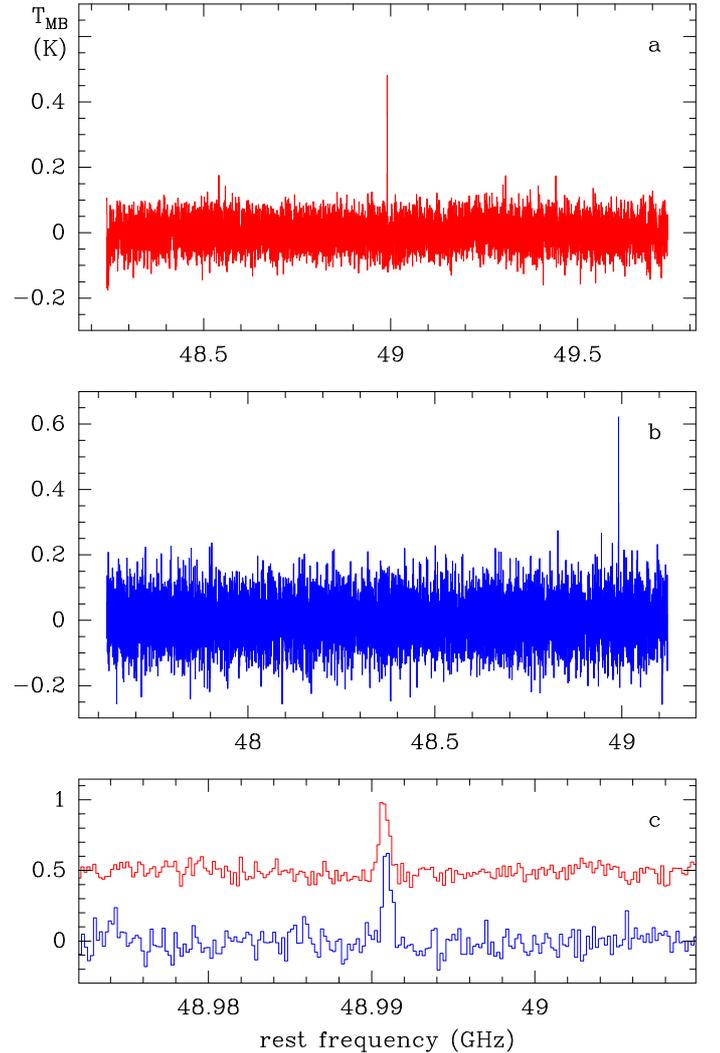}
  \caption{
CS $J=1\rightarrow0$ line observed toward W3(OH). ({\it a}) Line frequency 
in the band center. ({\it b}) Line frequency next to an edge of the band. 
({\it c}) A 40-MHz zoom of both spectra centered at the line. The comparison of 
both spectra is satisfactory, both in frequency and intensity scale. The upper 
spectrum (a) is less noisy than the lower one (b) owing to different integration 
times (8 and 3 minutes, respectively). 
  }
\end{figure}


A summary of the observations carried out in August and September 2011 is 
presented in Table 5. We used the Q-band receiver attached to the DSS-54 
antenna, and only the broad-band setup of the FFTS (setup 1 in Table 4), with 
just one FFTS card (e.g., a total instantaneous bandwidth of 1.5\,GHz). A 
second set of observations is planned to test the high-resolution modes, which 
will be reported as a Q-band catalog (Rizzo et al.~2012, in preparation).

Spectra of the observations reported here were reduced with simple procedures 
(e.g., baseline subtraction and FFT components editing) using CLASS. 
No spikes were noted when the input signal level to the FFTS was within 
the working regime (see Table 3); however, some spikes (up to four, depending on 
the observing frequency) were recorded when the signal was below these levels.
When possible, the spectra were compared with published data, provided in Table 
5. Some of the resulting spectra, after reduction, are shown in Figs.~4--8.

In Fig.~4 the CS $J=1\rightarrow0$ line toward W3(OH) is plotted. While in 
Fig.~4a the line is in the center of the band, it is close to a band edge in 
Fig.~4b. These observations were made to test the frequency and intensity 
stabilities in different parts of the band. As Fig.~4c shows, the comparison 
between both spectra are considered satisfactory.

Figure 5 shows a spectrum of the star-forming region W51D at 43\,GHz. Because 
this frequency may be tuned using both synthesizers, we used this case for 
cross-checking frequencies and intensities using different synthesizers (LO and 
HI in Figs.~5a and 5b, respectively) and during different days. System 
temperatures were similar during the three days, and therefore the noise level 
differences are due to different integration times (16 and 30 minutes in 
Figs.~5a and 5b, respectively). The combined spectrum, 1.9\,GHz width, is depicted 
in Fig.~5c. The spectrum is dominated by radio recombination lines (RRLs), the 
thermal ($v=0$) SiO $J=1\rightarrow0$ line, and its corresponding masers $v=1$ 
and $v=2$.

Figure 6 displays the well-known spectral wealth of Orion\,KL. The combined 
spectrum has 2.6\,GHz of bandwidth, and a total integration time of 45 minutes. 
As Fig.~6a shows, the spectrum is dominated by the thermal ($v=0$) and maser 
($v=1$ and $2$) emission of SiO $J=1\rightarrow0$. In Fig.~6b, some 
lower-intensity lines are noted, including the thermal isotopomer $^{29}$SiO 
$J=1\rightarrow0$, some RRLs, organic compounds, and the 44\,GHz methanol maser. 
The lower panels of Fig.~6 show a zoom of these and other lines, detailed in the 
caption. Part of this spectrum is reported for the first time. The strong SiO 
maser at 43.122\,GHz was used to study possible aliasing effects and to measure 
the image sideband rejection; no aliasing was measured, and the image band is 
attenuated by more than 6, 15, and 25\,dB at 25, 50, and 75\,MHz of the pass band 
borders, respectively.

Figure 7 depicts the spectrum toward Sgr\,B2(M), centered at 42.75\,GHz, and 
having 1.5\,GHz of bandwidth. A total of 24 minutes (on source) was employed. The 
full spectrum is shown in Fig.~7a, while in Figs.~7b-e a detailed zoom and line 
identification are displayed. As usual in this source, a mixture of emission and 
absorption features appear. Although no maser SiO lines are detected, the presence 
of the rare isotopomer $^{30}$SiO $J=1\rightarrow0$ line is remarkable, as well as 
several complex organic compounds and reactive ions. 

A broad composite spectrum toward TMC-1 is shown in Fig.~8. The bandwidth is 
8.5\,GHz, and the total integration time is 39 minutes; each 1.5\,GHz-wide 
individual scan was integrated for 3 to 8 minutes, as the different noise levels 
along the observed band show. More than a dozen lines are identified. In this 
source, the lines are detected just in a single channel, because the intrinsic 
linewidths are lower than 0.5 \kms\ \citep{tak98}, which is only a fraction of 
the channel width (1.2\,km\,s$^{-1}$) at Q-band frequencies.


\begin{figure*}
  \sidecaption
     \includegraphics[width=12cm]{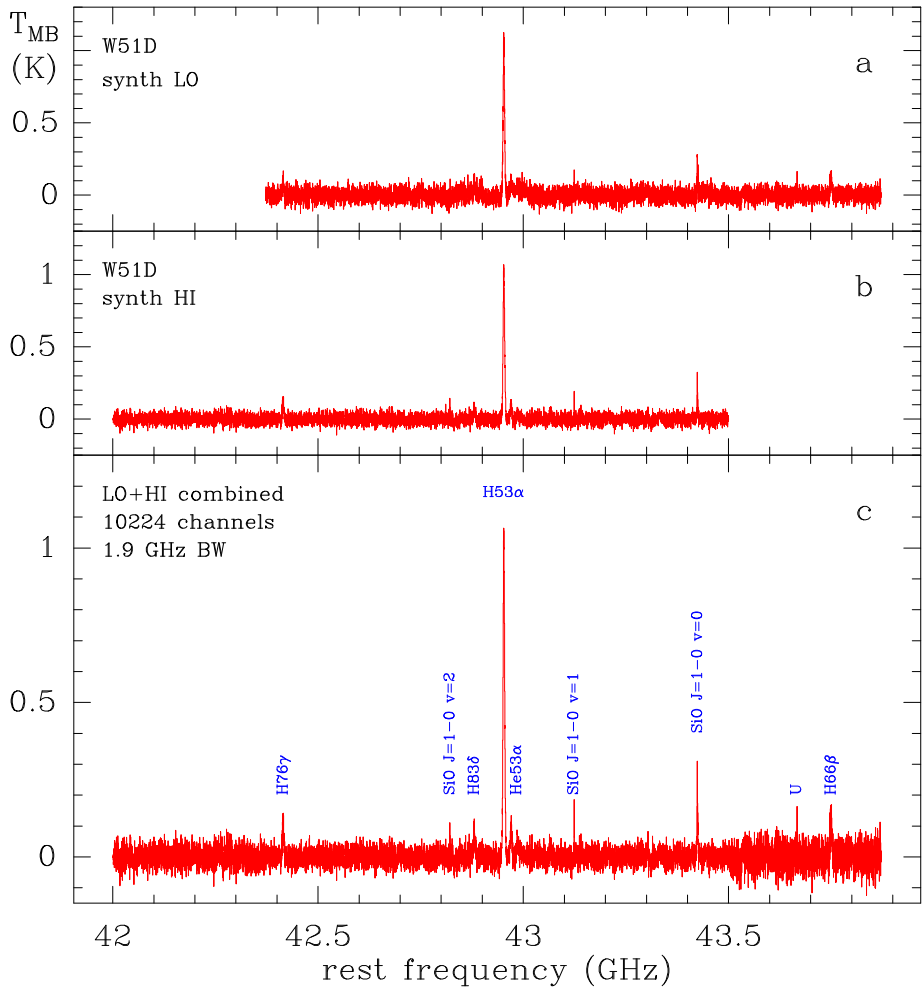}
  \caption
   {W51D at 43\,GHz. ({\it a}) Spectrum obtained using the synthesizer LO. 
   ({\it b}) The same, but using the synthesizer HI. ({\it c}) Combined spectrum, 
   covering a bandwidth of 1.9\,GHz. This test was made to check the consistency 
   of both synthesizers. Line identification is included in the bottom spectrum.}
\end{figure*}



\section{Improvements and upgrades}

As mentioned above, the commissioning tests reported here were made in the 
low-resolution mode (setup 1 in Table 4), and using just a single FFTS card. A 
second FFTS card and the high-resolution cores are planned for installation at 
the beginning of 2012. In low resolution, these new configurations will provide 
an instantaneous bandwidth of 3\,GHz, which would be connected to a single or 
both polarizations, and tuned at the same or different central frequencies. 
Furthermore, the system will provide high-resolution spectra (see Table 4), for 
up to two bands of 100 or 500\,MHz. Once these upgrades are completed, and the 
software is modified, we plan to start a second set of commissioning observations.

As part of the normal operation and maintenance activities, RF cables and 
cryogenics will be checked and replaced if necessary. The incorporation of new 
programmable attenuators and amplifiers adapted to the frequency range of the IF 
processor are also planned; these devices will maintain the signal within the 
working limits of the FFTS (Table 3) and will improve the flatness of the 
instantaneous pass band. New releases of SDAI will ease the simultaneous 
observations of different bandwidths and resolutions.

The system performance, in particular regarding possible harmonics and image 
sideband rejection, will be a subject of continuous attention. Finally, the backend 
can be made to be even more productive by the incorporation of two other FFTS cards, 
which will increase the instantaneous bandwidth to up to 6\,GHz. The IF processor 
and the SDAI software are already prepared for such an upgrade.

\section{Conclusions}

The facility reported here constitutes a significant enhancement to the radio 
astronomy capabilities at the MDSCC. The new backend performance fulfills the 
initial objectives. For the commissioning activities, we chose a set of sources 
based on line profile variety and spectral richness. The resulting spectra are 
considered to be of high quality. When the input signal to the FFTS fits within 
its working limits (Table 3), no spikes were noted and the baseline subtraction 
was performed without major inconveniences; ripples, when present, are easily 
removed by a FFT analysis. Weak RRLs are clearly distinguished from continuum 
after a few minutes of integration, and very narrow lines are easily recognized, 
although they appear in just a single channel.

This new facility improves the efficiency of the Robledo complex in two ways. 
On the one hand, simultaneous observations of several spectral lines in two 
polarizations are now possible, which makes better use of the available time, 
and improves the subsequent analysis by avoiding pointing and calibration 
problems. On the other hand, new scientific categories can now be addressed, 
such as spectral line surveys and extragalactic radio astronomy.

The incorporation of a second FFTS board and high-resolution cores are ongoing. 
First results using the high-resolution modes are planned, together with a 
Q-band catalog (Rizzo et al.~2012, in preparation), and possibly line surveys.

Finally, we point out that the project is in constant evolution, which means 
that the performance and reliability of the whole system is regularly checked 
and improved.

\begin{acknowledgements}
The backend was funded mostly through INTA grant 2009/PC0002CAB. This paper has 
been partially supported by MICINN under grant AYA2009-07304 and within the 
program CONSOLIDER INGENIO 2010, under grant “Molecular Astrophysics: The 
Herschel and ALMA Era – ASTROMOL” (ref.: CSD2009-00038). Part of this research 
was carried out at the Jet Propulsion Laboratory, California Institute of 
Technology, under a contract with the National Aeronautics and Space 
Administration. JRR wishes to thank Clemens Thum and Salvador S\'anchez (IRAM 
Granada) for their guidance in the first stages of the project. We are also 
indebted to Andreas Bell and Bernd Klein (MPIfR), and Ralf Henneberger (RPG), 
for their kind support about the FFTS and its libraries. We acknowledge our 
referee, Dr.~Jeff Mangum, for thoroughly reading the manuscript, and for 
comments and suggestions that allowed us to greatly improve this paper.
\end{acknowledgements}

%
\newpage
\begin{figure*}[b]
  \centering
     \includegraphics[angle=0,width=0.9\textwidth]{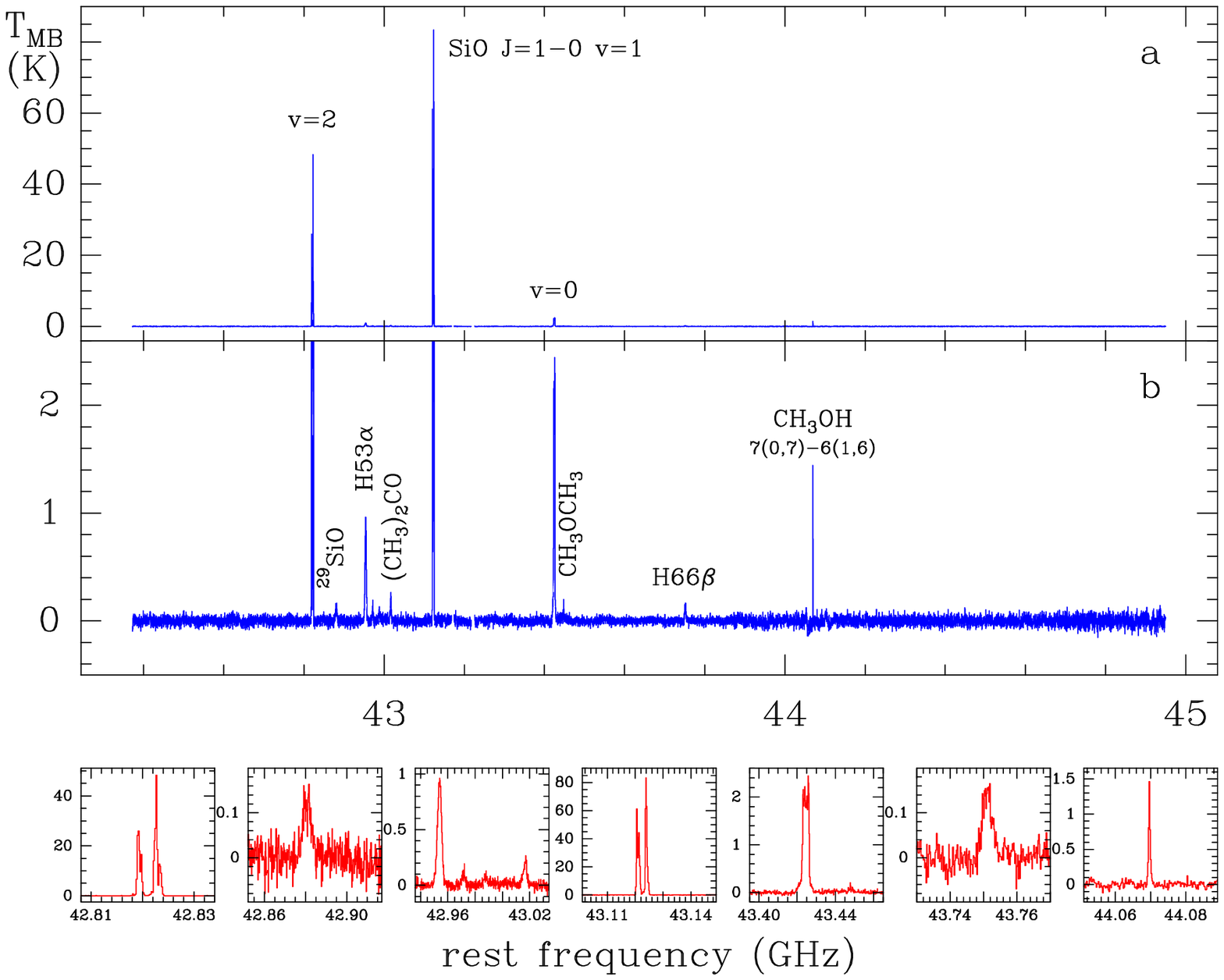}
  \caption
   {2.6\,GHz bandwidth spectrum of Orion\,KL.({\it a}) Full spectrum, 
    dominated by the SiO $J=1\rightarrow0$ lines, especially the vibrationally 
    excited states $v=1$ and $v=2$. ({\it b}) The same, but zoomed in intensity 
    to bring out other lines. ({\it lower panels}) Detailed view of some 
    spectral features. From left to right: SiO $J=1\rightarrow0$, $v=2$; 
    $^{29}$SiO $J=1\rightarrow0$, $v=0$, probably blended to H83$\delta$; 
    H53$\alpha$, He53$\alpha$, H89$\epsilon$, and (CH$_3$)$_2$CO 
    $10_{8,2}\rightarrow10_{7,4}$; SiO 
    $J=1\rightarrow0$, $v=1$; SiO $J=1\rightarrow0$, $v=0$, and a U line; 
    H66$\beta$; the 44\,GHz methanol maser.
   }
\end{figure*}
%
%

%
\newpage
\begin{figure*}
  \sidecaption
     \includegraphics[width=12cm]{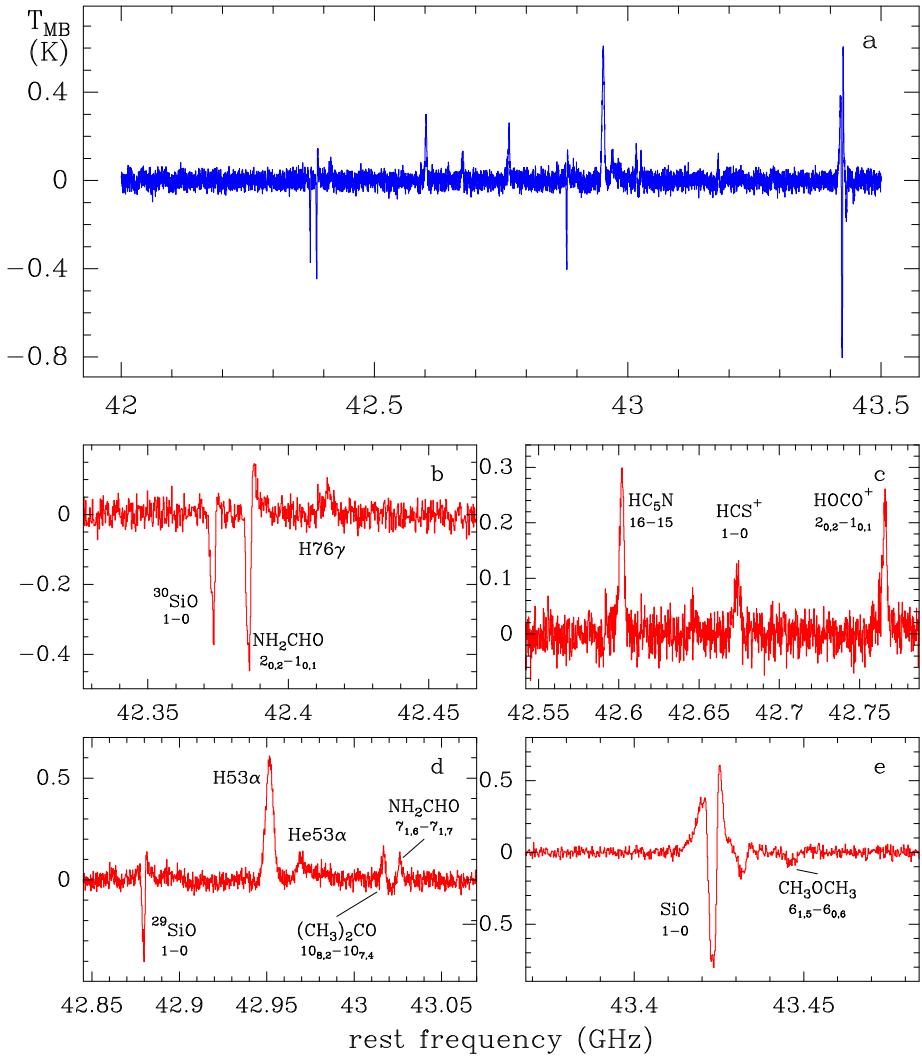}
  \caption
   {Spectrum of Sgr\,B2(M). ({\it a}) Full spectrum, showing the complexity 
    of this source. ({\it b-e}) Detailed view of some spectral features, 
    together with the corresponding identifications.
   }
\end{figure*}

\newpage
\begin{figure*}
\centering
  \includegraphics[width=0.9\textwidth,angle=0]{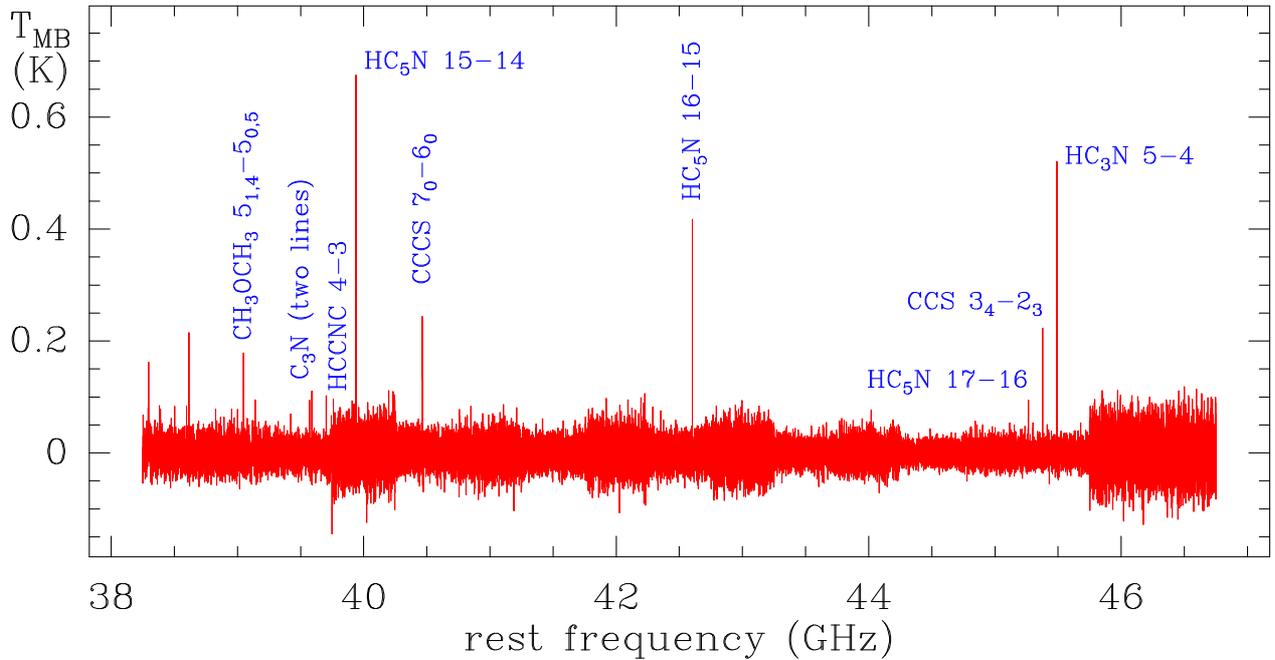}
  \caption{
   8.5\,GHz bandwidth spectrum toward TMC-1. The full spectrum is a 
   composite of individual scans of 1.5\,GHz of bandwidth. Integration time 
   of individual scans ranges from 3 to 8 minutes, which is easy to note by 
   different noise levels. Some of the identified lines are also indicated.
  }
\end{figure*}

\end{document}